\documentstyle[12pt,preprint,aps,tighten,epsfig,floats]{revtex}

\def\beq{\begin{equation}}
\def\eeq{\end{equation}}
\def\beqa{\begin{eqnarray}}
\def\eeqa{\end{eqnarray}}
\def\be{\begin{equation}}
\def\ee{\end{equation}}
\def\bea{\begin{eqnarray}}
\def\eea{\end{eqnarray}}
\begin{document}

\title{Spatial gradients in the cosmological constant}
\author{John F. Donoghue}
\address{Department of Physics and Astronomy, University of Massachusetts\\
Amherst, Massachusetts 01003 }
\maketitle
\thispagestyle{empty}
\setcounter{page}{0}
\begin{abstract}
It is possible that there may be differences in the 
fundamental physical parameters
from one side of 
the observed universe to the other. I show that the cosmological constant is
likely to be the most sensitive of the physical parameters to possible spatial
variation, because a small variation in any of the other parameters produces a huge 
variation of the cosmological constant. It therefore provides a very powerful
{\em indirect} evidence against spatial gradients or temporal 
variation in the other fundamental physical 
parameters, at least 40 orders of magnitude more powerful than direct experimental 
constraints. Moreover, a gradient 
may potentially appear in theories where the variability of the 
cosmological constant is 
connected to an anthropic selection mechanism, invoked to explain the smallness of 
this parameter. In the Hubble damping mechanism for anthropic selection, I calculate the
possible gradient. While this mechanism demonstrates the existence of this effect,
it is too small to be seen experimentally, except possibly if inflation happens around
the Planck scale.
\end{abstract}

\vspace{1.0in}

\section{Introduction}

Within the Standard Model of particle physics, describing our present 
best understanding 
of the fundamental forces and particles underlying all phenomena, there 
are 19 physical parameters or ``coupling constants''\cite{sm}. 
These include the 
fine structure constant $\alpha = 1/137$ as well as the masses of the quarks
and leptons. These are not predicted within the theory, but are
free parameters whose value must be extracted from experiment. However, within
the theory one feature is clear - these parameters are constants. They do not
depend on the time of the year nor our location in space. 

Despite this 
expectation, there exists a continual effort to look for spatial or temporal
variation in these parameters\cite{barrow}.  
It is certainly worthwhile to check this most basic
property of our physical theories. Moreover, as will be discussed
below, there exist physical mechanisms such
that different parts of the universe may have different values of the physical
parameters. These mechanisms have been invoked in cosmology
and in attempts to implement anthropic selection 
mechanisms\cite{anthropic,agrawal,weinberg,chaotic,freezing,gv2,jfd,bp}.
In these applications, the full Universe is 
much greater than the portion of it that we can observe. 
In addition there is the unusual feature is that the parameters
can vary, perhaps in a continuous fashion, in different regions.
More work is needed to construct
a fully realized cosmology incorporating these mechanisms, but it does appear 
physically possible to construct theories in which the parameters could
vary in different parts of the universe. 
At this stage, it is harder to know how we will test such theories. 
However, a possible test emerges if the variation of 
parameters is continuous. In this case, we may see a small variation 
when we look across our observable portion of the universe. Thus we can
look for a small continuous gradient in the parameters. 

I will argue that the cosmological constant is likely to be the parameter which
is most sensitive to spatial variations . As a consequence,
the cosmological constant provides extremely 
powerful indirect evidence against significant
variation of the other physical parameters, unless that variation occurs in several
parameters in a highly fined-tuned fashion.

I will also argue that a physical realization
of the anthropic principle\cite{anthropic} 
leads us to expect that there could be spatial variations
in the physical parameters and that such variation would be most sensitively
be manifest in the cosmological constant. While it is possible that the
effect may be too small to be observed, the opportunity to provide any form
of experimental evidence on the anthropic principle makes it imperative that this
variation be tested.  

\section{The cosmological constant and its potential variation}

The cosmological constant 
 $\Lambda$ is defined by the 
 vacuum value of the energy-momentum tensor $T_{\mu\nu}$, 
 $\mu,\nu = 0,1,2,3$ via
\beq
\langle 0| T_{\mu\nu} |0\rangle = \Lambda g_{\mu\nu}
\eeq
where $g_{\mu\nu}$ is the metric tensor. Since $T_{00}$ is 
the Hamiltonian, $\Lambda$ is equivalent to the energy density of 
the vacuum. It has dimensions of (mass)$^4$. The cosmological constant 
appears to be visible as an acceleration in the expansion of the
universe\cite{cosmological}.

There are many different contributions to the cosmological
constant. Each particle and each interaction contributes to
the vacuum energy.
In fact, each particle's contribution is infinite when calculated 
separately within the Standard Model because it diverges at high energy. 
However, the calculation of the high energy portion 
is unreliable since we do not yet know the correct theory that holds
at high energy. Even if the high energy contributions are cut off
by new particles or interactions (as happens for example in  
supersymmetry), contributions remain from those energies where we 
already understand the physics. Quantum Chromodynamics contributes 
effects of order 1 GeV$^4$. The electroweak theory contributes 
effects of order (100 GeV)$^4$. If supersymmetry is present in the 
theory but is broken at a low an  energy as is phenomenologically
possible, i.e. 1 TeV, then it yields contributions
of order (1 TeV)$^4$. If instead, the divergences are cut off at order
the Planck mass, $M_P=1.2 \times 10^{19}$~GeV, then the contributions
are of order $M_P^4$. 
 
The most important feature to be accommodated within
any theory of the cosmological constant is the extremely tiny
value of the final answer compared to the size of any of the
contributions.
The presently
favored experimental value ($\Omega_\Lambda \sim 0.7$) corresponds to 
\beqa
\Lambda_0 & = & 2.7 \times 10^{-47} ~{\rm GeV}^4   \nonumber \\
& = &  2.7 \times 10^{-59}  ~{\rm TeV}^4  \nonumber \\
& = &  1.2 \times 10^{-123} M_P^4
\eeqa  
The different contributions must 
nearly cancel to very many orders of magnitude. The situation is even
more difficult than if the experimental value were zero. In that 
case one might imagine that there exists a mechanism that forces the value to be 
exactly zero. 
However, with a non-zero experimental value, this hope must be
abandoned and we appear to be faced with {\it defacto} fine tuning of the 
different contributions with a tiny but non-zero residual.

If {\it any} of the contributions may have a spatial gradient, 
then the overall cosmological constant will also vary. Because of 
the large cancellation, a small percentage variation in a 
given contribution leads to a large percentage variation in 
the total.
A function that varies continuously in spacetime is a field, 
and so in this case at least some parameters are described by 
the values of one or more fields. There would continue to be contributions
to $\Lambda$ that are truly constant. Let us call these effects
$\Lambda_{\rm other}$.
If the mass scale for relevant 
underlying physics is $M_*$, these 
contributions to the cosmological
constant will be of order $\Lambda_{\rm other} \sim M_*^4$.
Added to this will be the contributions
due to the variable fields ($\Lambda_{\rm fields}$) 
which would also be of this order and 
which could occasionally (i.e. in some locations) cancel the other constant value,
\beq
\Lambda (x) = \Lambda_{\rm other} + \Lambda_{\rm fields}(x) 
\label{cancel}
\eeq
The picture that we are then led to involves these
fields having a range of values in the early universe such
that in some region the contributions largely cancel.  
The cosmological constant would not
be strictly a constant but would allow spatial gradients.
While locally the value could have little variation, a variation
could appear when comparing the value at different sides of the
observed universe.

Because of the enormous cancellation implied
in Eq.[\ref{cancel}] , a variation of 
$\Lambda_{\rm fields}$ of one part in
$10^{59}$ (assuming $M_* = 1$~TeV) would lead to a variation in
the net cosmological constant of order unity. Thus the search for
variations in $\Lambda$ provides an extremely sensitive probe for
variation in the underlying physics influencing $\Lambda_{\rm fields}$.

To our knowledge, there has been no phenomenological
investigation of a possible 
spatial gradient in the cosmological constant. A serious study of 
this possibility is warranted.

\section{Constraints on other parameters}

The search for a spatial variation in the cosmological constant is 
also an indirect search for the variation of other physical parameters.
This is because all the physical parameters influence the cosmological
constant in one way or another. Moreover, because the cosmological 
constant is so exquisitely sensitive, this indirect measurement is far
more powerful than direct searches. In this section we explore the
sensitivity of the cosmological constant in searches for variations of the 
other parameters. 

The possibility of spatial variations of the physical parameters has
been discussed by Barrow and O'Toole\cite{barrow}. 
At present there are suggestive but tentative indications 
for a temporal variation in the fine structure constant of order
\beq
{\Delta \alpha \over \alpha} \sim 10^{-5}
\eeq
from quasar spectral lines\cite{barrow2,temporal}. 
However, such a variation of $\alpha$ by itself would have an enormous effect
on $\Lambda$.
The fine structure constant influences the cosmological constant indirectly 
through its coupling to charged particles. 
Each charged particle receives a shift in its mass due to electromagnetism,
and the masses subsequently influence the zero-point energy. Therefore, changing
$\alpha$ changes the energy of the vacuum state. 

When calculated 
strictly within QED,
this change is infinite, since the electromagnetic self energy diverges. However, 
this divergence is only logarithmic, and hence is not deeply sensitive to 
the high energy scale of the complete final theory. We will proceed
in a fashion motivated by effective field theory. The divergence comes 
from high energy where we do not know the correct theory. A somewhat conservative 
approach is to calculate only that portion of the effect that comes from the
energy scales that we have already probed experimentally. This portion of the answer
will be unchanged by the introduction of new particles or interactions at high energy.
New physics will introduce new and different effects. While there appears to have
been a cancellation in the total magnitude of the cosmological
constant, this would not imply that there would be independent cancellations
in the variation of $\Lambda$ with respect to each of the basic parameters. 
We have no reason to expect 
that the net magnitude of the $\alpha$ dependence
will be smaller than the effects that we calculate 
here.
Therefore the estimate below can be treated as rough lower bounds on the sensitivity
of the cosmological constant to the variation in $\alpha$.

At present, we have 
the correct theory at least up to the scale of the weak interactions, which
can be taken to be approximately the Higgs vacuum expectation value
$v=246$~GeV. Let us therefore estimate
the effect of changing $\alpha$ due to radiative corrections below this scale.

The influence of the fine structure constant on $\Lambda$ can be estimated
via
\beq
{\delta\Lambda \over \delta\alpha} = \sum_{i}{\delta\Lambda \over \delta m_i}
{\delta m_i \over \delta\alpha}
\eeq
where the sum is over all particles. Both variations are greatest for the most
massive 
charged particle, which by far is the top quark ($m_t \sim 175$~GeV), 
so we can confine our investigation
to to that particle uniquely. In perturbation theory the variation of the top 
quark mass due to electromagnetic effect below the scale $v$ is given by the
self energy
\beq
\left. {\delta m_t \over \delta\alpha}\right|_{E<v}  = {m_t \over 2\pi}
\ln{{v^2 + m_t^2 \over m_t^2}}
\eeq 
Similarly the leading low energy contribution of the top quark to the 
vacuum energy is
\beq
\left. \Lambda\right|_{E<v}  = 2 \int^{v} {d^3k \over (2\pi)^3} {1\over 2} \omega_k
\eeq
where the extra factor of two is for the two different spin states.
This leads to an estimate
\beqa
\left. {\delta\Lambda \over \delta m_t}\right|_{E<v}  & = & 2\int^{v} {d^3k \over (2\pi)^3} {1\over 2} 
{m_t \over \omega_k}   \nonumber \\ 
& = & {m_t  \over 4\pi^2}\left( v \sqrt{v^2 + m_t^2} - m_t^2 \ln{{v  +
\sqrt{v^2 + m_t^2} \over m_t}}\right)
\eeqa
Combining these we get
\beqa
\left. {\delta\Lambda \over \delta \alpha}\right|_{E<v} 
& = & {m_t^2  \over 8\pi^3}\left( v \sqrt{v^2 + m_t^2} - m_t^2 \ln{{v  +
\sqrt{v^2 + m_t^2} \over m_t}}\right) \ln{{v^2 + m_t^2 \over m_t^2}} \\
& \sim & 10^{52} \Lambda_0
\eeqa
where $\Lambda_0$ is the present value of the cosmological constant 
quoted in Eq. 2. Note the extreme sensitivity - a fractional change in $\alpha$ 
of order $10^{-5}$, in isolation, would yield a cosmological constant 47 orders
of magnitude larger than allowed! Conversely, the lack of large variation
in the cosmological constant would constrain the variation in $\alpha$ in 
isolation to be 
less than less than one part in $10^{47}$. 

Similar estimates can be given for the effects of variation of other 
physical parameters. For example the same methods lead to the effect 
of varying the electron mass
\beqa
\left. {\delta\Lambda \over \delta m_e}\right|_{E<v}  
& = &  {m_e v^2 \over 8\pi^2} \nonumber \\
& = & 10^{42} {\Lambda_0 \over m_e}
\eeqa
and the Higgs vacuum expectation value itself,
\beqa
\left. v{\delta\Lambda \over \delta v}\right|_{E<v} 
& \approx &{m_t } \left. {\delta\Lambda \over \delta m_t}\right|_{E<v}     \nonumber \\
& = & 10^{52} {\Lambda_0 }
\eeqa
We see that all of these variations are highly constrained by even the roughest
lack of variation of the cosmological constant.

These estimates are in reality very crude since we are neglecting the physics
beyond the electroweak scale. This can easily be more important than the physics
which we do consider and could overwhelm the result presented above.
However, there is no reason to expect that physics 
from higher energy would lead to a smaller residual than that which we 
estimate\footnote{Recall that what is being studied here is not the overall 
cosmological constant but the the variation of $\Lambda$ with respect to the other
parameters. Even if we are faced with an apparent cancellation in the overall 
magnitude, it would take separate ``miracles'' if there were large cancellations
in the variation of $\Lambda$ with respect to each of the other parameters}. 
Given these caveats, it is important to treat the result not 
as a specific bound on the variation of $\alpha$, but as a qualitative 
lesson. We have learned that the variation of the fundamental parameters
can have a tremendous influence on the cosmological constant and that, unless
some special situation occurs, the information on the cosmological
constant already rules out the magnitudes of variation that can be tested in
direct searches.

The caveat within the previous sentence, about special situations, is needed
because there is a logical possibility that 
the variation of the parameters occurs in such a way that conspires to leave
the cosmological constant unchanged. If one considers the  
space of all possible values of the physical parameters (which in the
Standard Model is a 19 dimensional space), then there is a surface in that
space corresponding to a fixed cosmological constant. Along that surface
the variation due to one
of the parameters is compensated by the effects of changes in the other
parameters. It is logically possible that the only variations of the parameters
allowed are such correlated variations. For this reason, direct searches do give
independent information compared to the indirect constraints described
in this section. However, there is no known or suspected reason for allowing only
such correlated and fine-tuned variations.   

The previous considerations are also relevant for possible temporal variation of
the fundamental parameters. A time variation of the fine structure constant of the
order being discussed in the literature\cite{barrow2,temporal} would lead to an enormous 
variation of the cosmological constant. While it is possible that the 
cosmological constant was quite different in the extremely 
early universe, it could not have been
very different in the times that can be explored by direct observation through looking
at distant (older) objects. Significant
temporal
variation can only occur if other parameters are also variable in a highly fine-tuned
fashion. Note that although one accesses temporal variation by looking at
distant objects, it can be differentiated from spatial variation. A time
variability appears the same in all directions while a spatial variation would
be different in different directions.

\section{Anthropic gradient in the cosmological constant}

In this section, I argue that the mechanism which allow the anthropic selection
of a small cosmological constant may potentially leave a residual 
effect of a gradient in the cosmological constant. 

The anthropic principle\cite{anthropic} can be motivated by the observation that the
physical parameters are, fortunately, almost uniquely favorable for the 
existence of life. Very small changes in the parameters would lead to a
situation incapable of life. This can be true whether one uses a narrow definition
of life (``life as we know it''), or much broader definitions. As an example of the
first type, slight changes in the masses of the up and down quarks and/or electron
would lead to a world with no stable protons nor hydrogen atoms, which precludes
all organic chemistry as we know it. However, heavier atoms could still exist,
so perhaps a different form of life would be possible. In contrast, a variation
of a different sort, a slight increase of the Higgs vacuum expectation 
value (with the other parameters held fixed), would
lead to the instability of all elements except hydrogen\cite{agrawal} 
and would lack 
the complex chemistry which is likely
needed for {\em any} form of life. Similarly, a variation in the 
cosmological constant, small on its natural scale, would lead to a Universe where no
matter clumps gravitationally, which would again be sterile\cite{weinberg}. 
Given the potentially
allowed
range of the parameters in the Standard Model, the anthropically allowed region
is a very tiny part of the parameter space. This leads to the hypothesis that 
the observed values of the parameters are connected to the existence of life, which
is a general statement of the anthropic principle. 
There are physically reasonable situations where the anthropic condition is a 
natural constraint. This occurs, for example, if there are multiple regions in the 
universe that can have different values of the parameters. In this case, it is a
natural consequence that we would only find ourselves in one of those regions 
where the 
parameters were amenable for life.
The problem of the cosmological constant
is so severe that it motivates us to take seriously
a multiple
domain anthropic ``solution''\cite{weinberg,freezing,gv2,jfd}. 

A crucial question is whether the different values of the
cosmological constant vary discretely or continuously. 
Since we have a good understanding of the Standard Model
at present energies, a theory that leads to variable
parameters can only appear at higher energies which 
have not yet been investigated. Let us call the energy
scale of such a theory $M_*$, with $M_* \ge 1$~TeV. 
Consider the case when 
such a theory has a discrete set of ground states, each
leading to different values of at least some parameters.
Each of these ground states will have a cosmological
constant which is naively of order $M_*^4$ and the differences
between the values will also be of order $M_*^4$. 
In particular, 
the spacing between the smallest negative value of $\Lambda$
and the smallest positive value would then be expected
to be of this order. In this case,
it is extremely unlikely that any of the ground states has
a value of $\Lambda$ 
which falls close enough to zero to satisfy the
anthropic constraint. To require that the ground states
be so densely distributed near $\Lambda \sim 0$ that one has
a reasonable chance of satisfying the constraint would normally
require the existence of an extremely
small parameter 
within the theory. While this may be possible in very special 
cases\cite{bp}, it is normally unnatural that a theory
with discretely different parameters would be able to satisfy
the anthropic constraint in any of the domains. Therefore it is
plausible to consider potential theories with continuously variable 
parameters.
 
A theory with a continuously variable parameters could naturally
satisfy the anthropic constraint. The only requirement is that
both signs of the cosmological constant be possible. In this case,
there will be a region where the value goes through zero and 
in this neighborhood arbitrarily small values are possible. For this 
region to become our observed universe, we would need to invoke
an inflationary epoch in which the region greatly expands. However,
inflation \cite{inflation}
is likely needed in any case, so that this is not a
difficult additional requirement. In the early universe, gradients in the
parameters would naturally exist. Inflation would smooth these out, but 
some residual variation might exist. Thus we see that anthropic arguments 
concerning the cosmological constant also provide a motivation for a 
spatially varying $\Lambda$. 

There has been little work on mechanisms for variable parameters. Two mechanisms are
known, one involving scalar fields with very flat 
potentials\cite{freezing,gv2,jfd} and another
involving four-form fields\cite{gv2,jfd,bp} such as appear in string theory. I will
provide an estimate of the spatial variation in $\Lambda$ within the former mechanism.

 \section{Example: Hubble damping}

  Let us now study the particular mechanism corresponding
  to a light scalar field with a very flat potential. This was
  explored in more detail in ref \cite{freezing,jfd}. We will see that
  this mechanism necessarily does lead to fluctuations in the 
  cosmological constant across our visible universe, although the 
  magnitude of the effect will be small unless inflation takes place 
  close to the Planck scale.

  First we recall some results from previous work. The most
  important is that, in order to not be evolving significantly today, 
  the potential must be very flat, satisfying
  \beq
  V'(\phi) \le 10^{-122} M_P^3
  \eeq 
   where the prime denotes the derivative with respect to $\phi$. 
   Because this is so small, the field $\phi$ must itself be very large
   $\phi \ge 10^{58} M_P$ in order to influence the cosmological constant.
   In turn, the large value of the field can only be accomplished, while
   naturally keeping the gradient energy small, if there are extremely
   many, $N \ge 10^{148}$, e-foldings of inflation, such as occurs in 
   eternal inflation. This latter requirement comes from the fact that the 
   field $\phi$ exhibits a random walk in its magnitude during inflation, 
   with a quantum fluctuation of magnitude $\Delta \phi \sim H/2\pi$ in 
   a region of size $H^{-1}$ during each e-folding time of inflation\cite{fluctuations}. 
   These fluctuations
   allow $\phi$ to grow in magnitude in order to satisfy the 
constraint $\phi \ge 10^{58} M_P$ . 

It is the quantum fluctuations of $\phi$
that will also generate 
fluctuations in the cosmological
   constant. The logic is as follows. The key is that the  field $\phi$ has random 
fluctuations of size $H$ per e-folding time, $H^{-1}$, 
during the de Sitter phase of
inflation\cite{fluctuations}, 
but these are then effectively frozen in the subsequent radiation dominated
and matter dominated
phases after inflation. The regions that we see at opposite sides 
of the observable universe can have somewhat different values of the field $\phi$
because the field fluctuations have been randomly different during the inflationary
period. The field fluctuations in turn will lead to fluctuations in the cosmological 
constant.  
To estimate these fluctuations we must go through three
steps. First we must trace back our present universe to the time at the end of inflation
in order to find out how many different regions of size $H^{-1}$ there were in
the volume that led to our visible universe. Secondly, we need to see how different
the values of $\phi$ could have have been in those regions. We do that by tracing 
the regions back through the deSitter phase until regions from opposite sides of the
universe were in causal contact in a volume of size $H^{-3}$. This will allow us to 
estimate the number of e-foldings that occurred from the time of 
causal contact until the end of 
inflation, and hence will provide an estimate of how different the field values
can be in the different regions. The last and most uncertain step is to transform the
fluctuations in $\phi$ into fluctuations in $\Lambda$.
   
   The observable universe today consists of very many regions that
   were spatially disconnected at the end of the inflationary period. 
   Consider an initial region of size $H^{-1}$ at the end of inflation. 
   This will have expanded by a factor
   \beq
   {a_{now} \over a_{initial}} = 10^{32} {T_{r} \over M_P}
   \eeq   
   in the subsequent time. This formula emerges from tracing backwards the
   evolution of the scale factor from the present time
   into the epoch when it was 
   radiation dominated\cite{weinbergbook}. 
   Here $T_{r}$ is the initial reheating temperature 
   after inflation. With efficient conversion of the energy density 
   driving inflation into reheated matter we have 
  \beq
  H^2 \sim {T_r^4 \over M_P^2}
  \eeq
  Combining these factors, the initial patch has grown to a size
  \beq
  R_{now} \sim {a_{now} \over a_{initial}} H^{-1} \sim {10^{32} \over T_r}
  \eeq
  now. Despite the large growth this is still very small compared to the
  size of the universe
  \beq
  R_{universe} \sim 10^{26} {\rm meters} \sim {10^{61} \over M_P} \ \ .
  \eeq
  Thus the present universe consists of 
  \beq
  N_{regions} \sim \left({R_{universe} \over R_{now}}\right)^3 
  \sim 10^{87} \left( {T_r \over M_P}\right)^3  
  \eeq
  of these initial regions.
  
  Since the scalar field has been undergoing a random walk of 
  quantum fluctuations during the inflationary epoch, 
it will have a variation over these regions. 
Let us trace back the history of these domains in order to understand how
large the fluctuations in $\phi$ may be. To
produce 
  this number of domains from a single domain during inflation we
  need $N_{e}$ e-folding times, with 
 \beq
 e^{3N_{e}} = N_{regions}   \ \  \ .
 \eeq
 The resulting random walk in $\phi$ will produce
 a spread in values of order
 \beq
 \Delta \phi \sim \sqrt{N_{e}} {H \over 2\pi} 
 \eeq
 over the visible universe. 

While this will not be a uniform gradient, it 
 provides an estimate of the variation in $\phi$ over the universe.
In a multipole expansion a gradient term $\delta\phi_1 \sim x$ will be the leading
component. Since $\nabla^2 \delta\phi_1 = 0$, this component will be unaffected by the 
equations of motion and will survive until the present.
This is converted into a variation in the cosmological constant
 of size 
 \beq
 {\Delta \Lambda \over \Lambda } = 10^{123} {V'(\phi) \Delta \phi \over 
 M_P^4}
\le 10 \sqrt{N_{e}} {H \over 2\pi M_P}
 \eeq
Here is the weakest part of our estimate, as we have only an upper bound on
the derivative of the potential. However, this quantity cannot be too small
if this model is to be applied to anthropic ideas because if the
potential is too flat there will not be any influence on the cosmological constant.

Because we have only an upper bound on the variation of $\Lambda$ we cannot
give an unambiguous prediction. However, it is clear that under most 
circumstances the variation will be small. 
 If inflation occurs at the Planck scale it is possible that there could
be a visible effect, 
since we then have 
 \beq
 {\Delta \Lambda \over \Lambda } \le 10
 \eeq
 However, this bound drops rapidly with reheating temperature, resulting
 in 
 \beq
 {\Delta \Lambda \over \Lambda } \le 10^{-9}
 \eeq
 for $T_r = 10^{-5} M_P$. This mechanism does demonstrate the possibility
 of the variation in the cosmological constant, although its expected 
 magnitude is likely too small to be observed. The magnitude of the effect in 
other mechanisms remains to be explored.

\section{Conclusions}
 
We have seen that the cosmological constant is a very sensitive probe of 
any potential spatial variation within the underling theory. This can be used
to indirectly disfavor variations in the other parameters. However, the 
physics that
is employed in the anthropic solution to the cosmological constant problem
provides a novel motivation for searching for a variation in $\Lambda$. The
magnitude of the potential signal is presently unknown, and will likely be
small if the Hubble damping mechanism is correct. However, anthropic theories
will be in general very difficult to test and 
any possible signal of such an effect deserves serious investigation.

\section{Acknowledgements} 
I would like to that the theory group at CERN for its hospitality at the
start of this investigation and to thank Neal Katz, Martin Weinberg,
Jaume Garriga, Alex Vilenkin,
Norbert Straumann and Daniel Waldram for useful conversations. This 
work has been supported in part by the National Science Foundation and by
the John Templeton Foundation.

\end{document}